\documentclass{he_symp}
\usepackage{psfig}

\def\pdot {\dot P}

\def\ltsima{$\; \buildrel < \over \sim \;$}
\def\lsim{\lower.5ex\hbox{\ltsima}}
\def\gtsima{$\; \buildrel > \over \sim \;$}
\def\gsim{\lower.5ex\hbox{\gtsima}}
\def\msun{~M_{\odot}}

\def\uu {4U~0142$+$61}
\def\oo {1E~1048.1$-$5937}
\def\kes {1E~1841$-$045}
\def\axj {AX~J1845.0$-$0300}
\def\rx {1RXS~J170849$-$400910}
\def\rxs {1RXS~J1708$-$4009}
\def\ee {1E~2259$+$586}

\begin{document}
%
%

\title{The Anomalous X--ray Pulsars}
\author{S.Mereghetti\inst{1}, L.Chiarlone\inst{2,1}, G.L.Israel\inst{3} \and  L.Stella\inst{3}}
\institute{Istituto di Astrofisica Spaziale e Fisica Cosmica,
Sezione di Milano ``G.Occhialini'', CNR,
v.Bassini 15, I-20133 Milano, Italy
\and Universita' degli Studi di Milano, v.Celoria 16, I-20133 Milano, Italy
\and Osservatorio Astronomico di Roma, v.Frascati 33, I-00040 Monteporzio Catone, Italy }
\maketitle

\begin{abstract}
The Anomalous X-ray Pulsars (AXPs) are a small class of pulsars with spin periods in
the 6--12 s range,  very soft X--ray spectra, secular spin down on
time scales of $\sim10^{3}-10^{5}$ years, and lack of bright optical counterparts.
Two, possibly three, of them are close to  the centre of shell-like supernova remnants.
AXPs are one of the most enigmatic classes of galactic high-energy sources.
Isolated neutron stars powered by the loss of rotational energy can be
excluded on energetic grounds. The models based on neutron stars
involve either accretion (perhaps from a fossil disk around an isolated
neutron star) or the decay of a very strong magnetic field ($10^{14}-10^{15}$ G).
We review the models and the observational properties of AXPs,
including recent \textit{XMM-Newton} and \textit{Chandra} observations. We also present some unpublished
\textit{ASCA}  data.
\end{abstract}

\section{Introduction}

Bright X--ray sources powered by accretion in a binary system and showing regular
pulsations due to the characteristic lightûhouse effect from a rotating, magnetized
neutron star were discovered in the early seventies. More than one
hundred X--ray pulsars are currently known, most of  which are optically identified
with bright, massive early type stars, or provide other evidence of being in binary
systems.
The Anomalous X--ray Pulsars (AXPs) were recognized only in the last few years as
a separate class of objects, with
common characteristics clearly different from those of the normal pulsars.
Most remarkable is the   lack of signatures for a binary companion, a fact that prompted
a variety of  models based on isolated neutron stars or white dwarfs.

The interest for this small group (5 or 6) of pulsars has continued to increase in the
last few years due to several factors.
First, there is the exciting possibility that AXPs could be the  neutron stars with
the strongest known magnetic field, greater than several $10^{14}$~G.
Second, they share some similarities with the Soft Gamma-ray Repeaters
(SGRs, see Hurley 2000 for a review), another
class of neutron stars with exceptional properties including super-outbursts with
luminosity in excess of $10^{44}$ erg s$^{-1}$ (assuming isotropic emission).
Finally, it has become apparent that the ``\textit{textbook}'' example of association between
a neutron star and a supernova remnant, the Crab, is the exception rather than
the rule. Most supernova remnants do not contain ``\textit{Crab-like}'' pulsars, i.e.
neutron stars with high rotational energy loss,   magnetic field of several
10$^{12}$ G, small spin period, pulsed radio and non-thermal X/$\gamma$-ray emission,
and surrounded by bright radio/X-ray synchrotron nebulae.
This apparent lack of compact remnants is not considered a problem anymore,
since it has been recognized that there are different ways in which
young neutron stars manifest themselves. In this context, AXPs are relevant
since at least two of them (possibly three) are found right at the center of
shell-like supernova remnants.

\begin{table*}
      \caption{The AXPs and their   associated SNRs}
         \label{timing}
        \begin{tabular}{lcccccccc}
            \hline
            \noalign{\smallskip}
  SOURCE       & $P$  & $\pdot$~ & SNR  &  $P/2\pdot$ & B$^{(a)}$  & SNR~age & L$_{x}^{(b)}$ & D$^{(c)}$ \\
               & (s)  &  (s~s$^{-1}$)   & &    (years)  & (Gauss)    & (kyr) & (erg~s$^{-1}$) & (kpc) \\
            \noalign{\smallskip}
            \hline
            \noalign{\smallskip}
\ee    & 6.98    & $4.88\times10^{-13}$ & G~109.1--0.1&  227,000             & $6.1\times10^{13}$& 3 --17  & $10^{35}$ &  4 \\
      &         &                       & (CTB 109)  &   &         & & &  \\
\oo    & 6.45  &(1.5-4)$\times10^{-11}$&    --          & $(3-7)\times10^{3}$ & $3.2\times10^{14}$ & -- & $3.4\times10^{34}$ &  5 \\
  &   &   &      & &    &  & &   \\
\uu    & 8.69    & $1.98\times10^{-12}$&    --          & 70,000              & $1.4\times10^{14}$ & --  & $3.3\times10^{34}$ & 1 \\
 &  &   &    & &       &   & &   \\
\rxs   &11.00    & $1.9\times10^{-11}$ &    --          & 9,200               & $4.8\times10^{14}$ & -- &$6.8\times10^{35}$ &  8 \\
  &  &   &          & & &   & &   \\
\kes   &11.77    & $4.16\times10^{-11}$&  G 27.4+0.0       & 4,500               & $7.3\times10^{14}$ & $\lsim$3  & $2.3\times10^{35}$ & 7   \\
       &           &                   &   (Kes 73)  &  & &    & &   \\
\axj   & 6.97    & --                  & G~29.6+0.1    &  -- &  --    & $\lsim8$ &  $7.4\times10^{34 (d)}$&  8 \\
 \textit{(AXP candidate)}  &  & & & & & & &  \\
            \noalign{\smallskip}
             \hline
        \end{tabular}
\begin{list}{}{}
\item[$^{\rm (a)}$] magnetic field estimated from $B=3.3\times10^{19}~(P\pdot)^{1/2}$;
\item[$^{\rm (b)}$] luminosity in the 1-10 keV range, corrected for the absorption;
\item[$^{\rm (c)}$] assumed distance;
\item[$^{\rm (d)}$] assuming the blackbody model, see Table 2.
\end{list}
   \end{table*}


\section{Historical overview}

The AXP ``\textit{prototype}'' \ee\ was discovered at the center of the SNR G109.1--1.0
back  in 1980 (Fahlman \& Gregory 1981), and for many years it remained an isolated
oddity in the zoo of X--ray pulsars.
A small, positive period derivative of $5\times10^{-13}$ s s$^{-1}$,
similar to that of the Crab  pulsar, was measured (Koyama et al. 1987),
but it was clear that, owing to its long spin period (7 s), the loss of
rotational energy was orders of magnitude too small to power the observed
luminosity of a few 10$^{35}$ erg s$^{-1}$.
Thus \ee\ was not a rotation powered Crab-like pulsar and it also seemed different
from the majority of accreting pulsars known at the time, due to its soft X-ray
spectrum,  lack of a bright (and hence massive) companion star,
and   secular spin-down.

The discovery of several new X--ray pulsars, as well as the extensive efforts to
identify and study their optical counterparts, led to the following picture
by the end of 1994: out of a total of  39 accreting  pulsars known, only
three were in Low Mass X--ray Binaries (LMXRBs) with an optically
identified companion star: GX 1+4 (with an M giant companion of mass
M$_{c}\sim0.8-2\msun$), 4U 1626--67 (M$_{c}\sim0.02$ or $0.08\msun$),
and Her X--1 (M$_{c}\sim2\msun$). In addition there were four
pulsars, without   optical identification,
but with   optical limits excluding a massive companion star (\ee, \uu, \oo, RX
J1838.4--0301).  Mereghetti \& Stella (1995) pointed out the similarities in their
properties and the remarkable fact that they all have spin periods in the narrow interval
between 5 and 9 s.  For comparison, the High Mass X--Ray Binary
(HMXRB) pulsars known at that time had  periods
from 69 ms to 1450 s: a range spanning  more than four decades.
Mereghetti \& Stella (1995)
suggested that, together with 4U 1626--67 (P=7.7 s), these objects belong to a
different class of pulsators,  probably powered by accretion from a very low mass
companion  (this was known to be true at least for 4U 1626--67 which has an orbital
period of   41 minutes).

Van Paradijs et al. (1995) interpreted these
``\textit{6-sec pulsars}''\footnote{the name Anomalous X--ray Pulsars
was introduced later, see Duncan (2002)}
as  isolated neutron
stars accreting from a residual disk, of course with the exception of  4U 1626--67,
which, however, could be considered a close relative in the context of the evolutionary
model of Ghosh et al. (1997).

A few years later,   pulsations at periods similar to those of AXPs
were discovered in the quiescent X--ray counterparts of two
SGRs  (P=7.5  and 5.2 s, respectively in SGR 1806--20 and SGR 1900+14,
Kouveliotou et al. 1998, Hurley et al. 1999a). These sources were found to spin-down
with  period derivatives  $\pdot\sim8$ and 6 $\times10^{-11}$ s s$^{-1}$,
as expected for   dipole magnetic braking  in the  ``\textit{magnetar}'' model,
developed by Duncan \& Thompson to explain the SGR phenomenon.
This led, by analogy, to apply the same model also to AXPs (see Section 4.2).

In the meantime, the original list of AXPs had changed, not only for the exclusion
of 4U 1626--67, but also because it was found that, contrary to previous claims
(Schwentker 1994),
RX J1838.4--0301 was not a pulsar (Mereghetti et al. 1997).
These losses were balanced by the discovery of
two new pulsars with all the characteristics to be included in the
AXP group: \kes, discovered by Vasisht \& Gotthelf (1997) inside the Kes 73 SNR, and \rx\
(Sugizaki et al. 1997).
The last addition to the AXP group is possibly \axj\ (Torii et al. 1998),
but for the reasons described below, this should be considered only  a ``\textit{candidate}'' AXP.
Table 1 gives the  list of currently known AXPs.

\section{Main properties}

The main  characteristics  that distinguish AXPs from the more common
HMXRB pulsars are the following:

a) The absence of a massive companion star in AXPs has been established by means of
deep optical and IR observations. Even for the AXPs with the highest interstellar
reddening, the current data are sensitive enough to rule out the presence of an early type
mass donor. Further evidence comes from  the lack of  orbital motion signatures, such as
periodic intensity variations (eclipses   or dips) and pulse arrival times delays
(however this   analysis has not been carried out yet for all AXPs).

b) The AXPs are characterized by soft X--ray spectra, clearly different from those of
HMXRB pulsars. The latter have relatively hard spectra, usually well described by a
power law with photon index $\alpha_{ph}\sim$1 and an exponential
cut-off above $\sim$20 keV. In contrast, AXP spectra require much steeper power laws
($\alpha_{ph}>$3) and/or blackbody models with low temperature (kT\lsim0.5 keV).

c) In general, accreting neutron stars tend to spin-up secularly, due to the angular
momentum transferred from the accreting material.
Indeed this is observed in many HMXRB pulsars in which there is evidence for an
accretion disk. Wind-fed pulsars often show alternating episodes of spin-up and spin-down.
on the contrary, the spin periods of
AXPs are monotonically increasing at a nearly constant rate
(on timescales ranging from $\sim$5,000
to $\sim2\times10^5$ yrs).

d) Spin period in the range 5--12 s.
It is interesting to note that all the pulsars that satisfy the three above
criteria (a), (b) and (c) have periods in this very narrow range.

e) Rather constant luminosity, at a level of  $\sim$10$^{34}$û-10$^{36}$ erg s$^{-1}$.
In general, AXPs have
relatively steady X--ray fluxes, compared with the kind of variability displayed by
other classes of accreting compact objects. As described in
Section 5.4, \axj\ showed a flux decrease greater than a factor 14
in observations spaced 3.5 years apart.
If further  observations will confirm this source as an AXP, the interesting
possibility that a class of \textit{transient} AXPs exist should be considered.

f) Two AXPs are clearly associated to SNRs, three if also \axj\ is considered (see
Section 8.1).

To summarize, one might define an AXP as ``\textit{a spinning down  pulsar, with a soft
X--ray spectrum, apparently not powered by accretion from a companion star, and with a
luminosity larger than the available rotational energy loss of a neutron star}''.

\section{Models}

\subsection{Accretion based models}

Accretion is a well known powering mechanism that might be  at work also for AXPs.
Three different reservoirs  for the accreting material have been considered:
a companion star, the interstellar medium and a residual disk.

The first possibility was originally proposed
by Mereghetti \& Stella (1995), who suggested that the
AXPs were neutron stars with  very low mass companions, and
characterized by lower luminosities and higher magnetic fields
($B$$\sim$10$^{11}$\,G) than  classical LMXRBs.
However, no   detailed binary model for AXPs
has been developed yet. Indeed, the lack of evidence for a
companion star  stimulated more interest in models based on
isolated compact objects.

Accretion from the interstellar medium (ISM)  was suggested for \uu,
based on its spatial coincidence with a nearby  molecular cloud (Israel et al. 1994).
However, it is   unlikely  that this can explain all AXPs,
since ISM accretion can only provide a   luminosity

$$ L_{acc}\sim10^{32}~~v_{50}^{-3}~~n_{100}~~~~~\textrm{erg~s}^{-1} $$

\noindent
where  v$_{50}$ is the relative velocity between the
neutron star and the ISM in units of 50 km s$^{-1}$ and
n$_{100}$ is the gas density in units of 100 atoms cm$^{-3}$.
Even for extreme values of v$_{50}$  and n$_{100}$, this is clearly too small a luminosity,
unless all AXPs are   nearby  objects ($\sim$100 pc), a possibility that
seems very unlikely considering
their low galactic latitude,   high
X--ray  absorption, and (in at least two cases) the
distances of  the associated SNRs.

For these reasons, the most promising  models are based on ``\textit{fossil}'' accretion disks
(of different origin) around isolated neutron stars.

The scenario of an isolated  neutron star fed from a residual   disk
was first advanced by Corbet  et al. (1995) for \ee.
van Paradijs et al. (1995)  proposed that AXPs could be one possible outcome of
the common envelope evolution of close HMXRBs, in which
a residual accretion  disk is
formed after the  complete spiral-in of the neutron star
in the envelope of a giant companion
(a so called Thorne-Zytkow object, TZO, Thorne \& Zytkow 1977).

According to Ghosh et al. (1997),
a HMXRB undergoing common envelope evolution can produce two kinds
of objects, depending on the (poorly known) efficiency with which the
envelope of the massive star is lost.
Relatively wide systems are expected to have enough orbital energy to lead to
the complete expulsion of the envelope before
the neutron star settles at the center of the massive companion.
This   results in the
formation of tight binaries composed of a neutron star and a helium star,
like 4U~1626--67 and Cyg X--3.
Closer HMXRB, on the other hand,   produce TZOs, as a result of
complete spiral in of the neutron star in the common envelope
phase, and then evolve into AXPs.

Another possibility for the formation of a disk
is through fallback of   material from the
progenitor star after the supernova explosion.
For appropriate values of the neutron star magnetic field,
initial spin period, and mass of the residual disk (Chatterjee et al. 2000),
these systems can evolve into AXPs with   luminosities, periods and lifetimes
consistent with the observed values.
Due to the steadily declining mass accretion rate,
the rotating neutron star evolves through different states.
During an initial ``\textit{propeller}'' phase, lasting a few thousand years, the
spin period increases up to values close to those observed in AXPs.
In this phase, the AXP progenitors are  very faint, undetectable X--ray sources,
since accretion down to the neutron star surface is inhibited
(or greatly reduced) by the magnetospheric centrifugal barrier.
In the following phase, the spin frequency approaches the Keplerian
frequency at the inner edge of the disk $\Omega$(r$_m$), most of the mass flow is
accreted, and the star becomes visible as an AXP. During this
quasi-equilibrium phase, the neutron star  spins down trying to
match  $\Omega$(r$_m$), which decreases with the declining mass accretion rate in
the disk.

Accretion from a fossil disk has also been considered by Alpar (2001) and by
Marsden et al. (2001).
The latter authors (see also Rothschild et al. 2002) claim that SGRs and AXPs
are born in dense regions of the interstellar medium,
while the core collapse supernova explosions producing normal
neutron stars occur preferentially in superbubbles of hot and tenuous gas
surrounding the OB associations.
In their model, the difference in the environment, rather than intrinsic neutron star properties,
are at the origin of the AXP/SGR peculiarities by favoring the disk formation.
This could happen in two ways.
In the first one, the dense surrounding gas confines the SN progenitor
wind, thus favoring a reverse shock in the expanding remnants and the
formation of a ``\textit{push-back}'' disk (Truelove \& McKee 1999). The other way applies
in the case of a high velocity neutron star which can more easily capture part of the
nearly comoving   ejecta slowed down by the prompt reverse shock expected
in a dense environment.

This evidence for environmental differences in AXPs and SGRs is
based on the uncertain distances and ages of  SNRs
and has been criticized by Duncan (2002). Furthermore, some of the associations with
SNRs  considered by Marsden et al. (2001) are very unlikely to be true.

\subsection{Magnetar models}

Magnetars are neutron stars in which the dominant source of free energy
is provided by a strong magnetic field, rather than   rotation (as in the ordinary
radio pulsars).
They are supposed to have internal fields higher, by a factor ten ore more, than
the quantum critical value $B_{c} = \frac{m_{e}^{2}c^{3}}{e\hbar}=4.4\times10^{13}$~G
(at $B>B_{c}$ the energy between the Landau levels of electrons is greater than their
rest mass).

A magnetic field in this range can be produced by an efficient dynamo mechanism that
operates if the neutron star is born with a very short ($\sim$1 ms)
period  (Duncan \& Thompson 1992).
Although no direct observational evidence for such short initial periods has been found
in radio pulsars or in young supernova remnants like  SN 1987A,
Thompson \& Duncan (1995) pointed out that, in the case of the SGRs, a high magnetic
field is motivated by several independent requirements
(some of which are
based on the energetic and spectral properties of the
giant flare of 1979 March 5 from   SGR 0525--66).
Among the different pieces of evidence for a high B field is
the magnetic confinement of the hot pair plasma responsible for the
long soft tail of the flare and  the reduction in photon opacity required to exceed by
a factor  $\gsim10^{3}$     the Eddington limit in the soft $\gamma$-ray bursts
(an argument first applied to SGR 0525--66 by Paczynski 1992).
Furthermore, a strong dipole field is required to slow-down the neutron star   to
the presently long period (8 s) within a time ($\sim10^{4}$ yrs) compatible with
the age of the associated SNR.

In the magnetar model applied to the SGRs the magnetic field
is the main energy source, powering both the normal bursting activity,
characterized by     brief ($<$ 1 s) and relatively soft (peak photon energy
$\sim$20-30 keV) bursts of super-Eddington luminosity,
and the much more energetic flares that have  so far been observed only on
two occasions: on March 5, 1979 for SGR 0526-66 (Mazets et al. 1979, Cline et al. 1980),
and on August 27, 1998 for SGR 1900+14 (Hurley et al. 1999b, Feroci et al. 2001).
The diffusion of the magnetic field through the neutron star core causes fractures
in the crust that can inject energy in the magnetosphere producing the soft $\gamma$--ray
bursts, while large-scale magnetic reconnections are responsible for the exceptionally
intense flares (see Thompson (2001) for an extensive review).

The discovery of $P$ and the   measurement of $\pdot$ in SGR 1806--20
(Kouveliotou et al. 1998)
and  SGR~1900+14 (Hurley et al. 1999a, Kouveliotou et al. 1999) is generally
considered as a confirmation of the existence of magnetars.
In fact, interpreting the spin-down  as due
to magnetic dipole radiation losses, the
neutron star magnetic field can be derived as
B $\sim$ 3.3$\times$10$^{19}$ ($P\pdot$)$^{1/2}$ G,
yielding B $\sim$8$\times$10$^{14}$ G and $\sim$6$\times$10$^{14}$ G,
respectively for SGR 1806--20 and SGR 1900+14.
A more realistic estimate should take into account other contributions
to the spin-down torque, such as a particle wind outflow (Thompson \& Blaes 1998),
either continuous or in the form of sporadic   outbursts.
Lower magnetic fields are obtained when these effects are considered
(Harding et al. 1999).

As mentioned above, other arguments supporting the magnetar scenario are related
to the apparent youth of the SGRs, as inferred from their association to SNRs.
This is a matter of debate, since, while the   SGR 0526--66 / N49 association
is generally regarded as very likely, the same is not true for the other proposed
associations: SGR 1806--20 / G10.0--0.3 and SGR 1900+14 / G42.8+0.6 (see Gaensler et al. 2001,
Duncan 2002).

The magnetar model is very successful to reproduce those
temporal and spectral properties of   SGRs that are difficult to explain
with alternative models like  sudden accretion events
(see Thompson \& Duncan 1996).

The interpretation of AXPs in terms of magnetars is   a fairly  natural consequence
of the analogies between these two classes of objects.
Indeed, some similarities  between the prototype AXP \ee\ and  SGR 0526--66,
for which pulsations at 8 s were observed during the 5 March 1979 giant flare,
were immediately pointed out.
AXPs and SGRs are similar for what concerns their $P$ and $\pdot$ values
(and consequently the B inferred from the dipole braking formula), the
X-ray luminosity ($\sim10^{34}-10^{36}$ erg s$^{-1}$ for SGRs in quiescence,
Mereghetti et al. (2000)), and the association  with SNRs (at least in some cases).

The enormous  magnetic energy available in a magnetar can easily  power the
steady X--ray luminosity observed in AXPs. The magnetic field decay, on a
timescale of $\sim10^{4}$ yrs, heats the neutron star surface that emits thermal
radiation in the X--ray band (Thompson \& Duncan 1996). In addition, the multiple
small scale fractures induced by the magnetic field on the neutron star crust
produce an emission of Alfv$\acute{e}$n waves in the magnetosphere that accelerate
particles and can produce non-thermal emission.

Heyl \& Hernquist (1998) showed that,
if   B is greater than  $\sim10^{15}$ G, the residual thermal
energy of the neutron star is sufficient to power for a few thousand years
the observed X--ray luminosity, even without resorting to field decay.
This requires the presence of an  envelope of
hydrogen and helium (an iron envelope is much more
efficient in insulating  the core, resulting in a
lower luminosity and effective temperature at the
neutron star surface). The envelope of light
elements, with a   mass of $\sim$10$^{-11}$--10$^{-8}$ $\msun$,
could be due to fall-back of material after the supernova
explosion and/or   accretion from the interstellar medium,
if the neutron star were born in a sufficiently dense environment
($\gsim$10$^4$ cm$^{-3}$).

Which are the predictions of the magnetar model for AXPs?
In the simplest interpretation one would expect  a more stable
luminosity, spectrum and $\pdot$ noise,
than in accretion-powered pulsars.
The spectra and light curves expected from the surface of highly magnetized
neutron stars have been computed by several authors
(\"{O}zel 2001, Zane et al. 2001, Ho \& Lai 2001).
As described in Section  6.1, these can reproduce the salient observational
properties of AXPs. No detailed predictions exist yet for what concerns the
emission at other wavelengths. Accurate timing observations were, somewhat naively,
thought to provide a clear discrimination in favor or against the magnetar hypothesis,
but, as described in Section 6.2, the situation is not as simple.
In conclusion, we believe that the AXPs could be magnetars, but we are still far
from a decisive proof, and valid alternative models cannot be ruled out as easily
as it has been done for the SGRs.

\subsection{Other  models}

The momentum of inertia of  a white dwarf is a factor $\sim$10$^{5}$ larger than that of a
neutron star, leading, to a rotational energy loss $\dot{E}_{rot}$
greater than the AXPs luminosity:

 $$\dot{E}_{rot}=I_{wd}\Omega\dot{\Omega}\sim
 4\times10^{37}\left(\frac{P}{10~s}\right)^{-3}\
 \left(\frac{\pdot}{10^{-11} s~s^{-1}}\right)  \textrm{erg~s}^{-1}$$

\noindent
This led  Paczynski (1990) to propose that \ee\ could be a rotation-powered,
isolated white dwarf (following an earlier suggestion by Morini et al. (1988)).
In order to rotate with such a short period without breaking-up,
the white dwarf should have a mass greater than $\sim$1.3$\msun$.
Paczynski suggested that a fastly spinning, highly magnetized (B$>10^{8}$ G) and
massive white dwarf could be the result of a recent merger of two white dwarfs.
According to this model the AXPs X--ray emission could result either from
non-thermal processes powered by rotational energy loss, as in young radio pulsars,
or from the thermal emission of a hot white dwarf.

It is not clear whether this model, originally developed for \ee\, could
account for the formation of a SNRs, since the outcome of a white dwarfs merger
is highly uncertain. Another problem is that no large $\pdot$ variations are expected
owing to the large momentum of inertia of a white dwarf.

A different scenario for AXPs (and SGRs), based on strange matter stars, has
been proposed by Dar \& DeR\'{u}jula (2000). According to these authors the AXPs
are either strange stars or quark stars in which the X--rays  are powered by
gravitational contraction and the spin-down is due to the emission of relativistic jets.

\section{X-ray spectral properties}

\subsection{Phase--averaged X--ray spectra}

AXPs have very soft spectra.  Early data of low statistical
quality were sufficiently well fit by
steep power laws yielding $\alpha_{ph}\gsim3$ (while single blackbody
models were unable to reproduce the observed flux above a few keV).
When better observations became available with \textit{ASCA} and, especially,
\textit{BeppoSAX} (the latter extending to energies as low as 0.1 keV),
it became clear that a more complex spectrum, consisting
of a blackbody-like component
with kT$\sim$0.5 and a steep power law ($\alpha_{ph}\sim$3--4), is required.
This spectral decomposition has become the \textit{canonical} AXP spectrum
(see the spectral parameters in Table 2),
but it should be noted that equivalently good fits can be obtained, e.g., with
the sum of two blackbody components (Israel et al. 2002).

It is likely  that, due to current instrumental limitations,
the canonical two-component model is
an oversimplified description of the true underlying spectral continuum.
However, it is tempting to interpret the two spectral components
in terms of two   distinct
processes and/or emitting regions in order to infer some physical
parameter.
For example, the blackbody-like  component can be naturally interpreted as
thermal emission leading to  emitting areas of the order
of a few km$^2$. These are smaller than the whole neutron star surface,
but yet much greater than the polar cap hot spots expected in case of magnetically
channeled accretion.

Perna et al. (2001) showed that larger emitting areas, consistent with
the dimensions of a neutron star, can be obtained
with a model for the thermal emission from a light-element
atmosphere with magnetic field B$\gsim$10$^{12}$ G.
Such spectra are typically harder than the corresponding blackbodies,
thus giving best fits with lower temperatures and higher normalizations.
Interestingly,  the inclusion of a separate power law component is still required.

There are no clear correlations between the spectral parameters of the canonical
model, but some spectral properties have been found to correlate with other quantities.
For example, Marsden \& White (2001) found that the AXPs with the highest spin-down
rate have the hardest spectra, while
Israel et al. (2002) noted a correlation between the blackbody contribution to
the total luminosity, $L_{BB}/L_{tot}$, and the pulsed fraction (see Fig.1).

In interpreting correlations of this kind one should always
remember that, as mentioned above, there is no compelling evidence
for the actual presence of two physically distinct emitting components.
In fact, as noted by \"{O}zel et al. (2001),
the blackbody contribution  to the
flux is a strong function of  the energy, while the pulsed
fraction at different energies does not change significantly.
In the case of two physically distinct spectral components
this would require a rather unlikely a ``{\it ad hoc}'' coupling  in the pulse profiles of the
blackbody and power law components.

\begin{figure}
\centerline{
\psfig{file=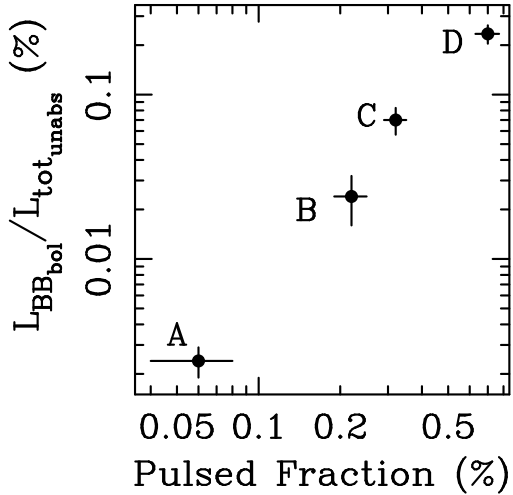,height=7cm,angle=-90} }
\caption{Correlation between $L_{BB}/L_{tot}$ and the pulsed fraction.
$L_{BB}$ is the bolometric luminosity of the blackbody component,
$L_{tot}$ is corrected for the absorption and refers to the 0.1--10 keV range.
A=\uu, B=\ee, C=\rxs, D=\oo\ (Israel et al. 2002).
\label{corr}}
\end{figure}

\subsection{Phase resolved X--ray spectra}

Spectral variations as a function of the pulse phase are present in practically all
kinds of pulsars and AXPs make no exception, as can be seen at least
in some of  the light curves
of Fig.2  In general these variations are not large enough (or the data lack
sufficient statistical quality) to be characterized as significant variations
in  the spectral parameters derived from the fits (see, e.g., Parmar et al. 1998,
Patel et al. 2001, Israel et al. 1999a, Oosterbroek et al. 1998).
The best evidence for phase-dependent spectral parameters was found in  \rxs,
where Israel et al. (2001) measured a variation of the power law photon index
from 3.1 to 1.9 across the pulse (see also Gavriil \& Kaspi 2002).
In contrast, \oo\  shows a very energy-independent pulse profile (see Fig.2).
However, a recent \textit{XMM-Newton} observation of this source revealed that a ``\textit{soft
excess}''  with respect to the phase-averaged spectrum is present below $\sim$1.5 keV
in correspondence of the pulse minimum (Tiengo et al. 2002).

\subsection{Narrow spectral features}

Cyclotron resonance features would provide  useful diagnostics,
but so far no convincing detection has been reported.
The energy of the line is

$$ E_{Cycl}~ \approx ~11.6~ Z_g~ \left( B \over 10^{12}~ G \right)~~ \textrm{keV} $$

\noindent
for electrons, and

$$ E_{Cycl}~ \approx ~0.63~ Z_g~ \left( B \over 10^{14}~ G \right)~~ \textrm{keV} $$

\noindent
for protons, where $Z_g ~=~ \sqrt{1 ~-~ 2GM/Rc^2}~~$ is the gravitational redshift
at the neutron star surface,
typically in the range 0.7--0.85.
Model atmosphere calculations (Zane et al. 2001) predict that proton cyclotron features in
highly magnetized neutron stars
are conspicuous (EW up to $\sim 100$~eV) and relatively broad
($\Delta$E/$E_{Cycl}$ $\sim$ 0.05-0.2).
However it is  clear that the discovery of \textit{a single} line would
not permit to easily distinguish between the electron and ion case and  thus
establish the magnetic field strength in an unambiguous way.

Early reports of a pulse phase-dependent line in \ee\
(Koyama et al. 1989, Iwasawa et al. 1992) were not confirmed
by subsequent higher quality  observations.
Sensitive line searches with the \textit{XMM-Newton} and \textit{Chandra} satellites
have been performed so far only for three AXPs:
Tiengo et al. (2002) reported the possible presence of an absorption
line at 4.71 keV in \oo\ during a short (5 ks) \textit{XMM-Newton} observation,
but more data are required to confirm this line (note  that it was seen only in one of the
three EPIC CCD cameras). \uu\ was observed with the \textit{Chandra} High Energy Transmission
Grating Spectrometer by Juett et al. (2002).
For  absorption lines in the range $\sim$0.7--5 keV, they reported upper limits
from a few eV to a few hundred eV
(depending on energy and assumed width of the line).
No significant spectral features were found in the range 0.5--7 keV  by
Patel et al. (2001) who observed \ee\ with the \textit{Chandra} ACIS instrument.

\begin{table*}
      \caption{Spectral parameters and  fluxes of AXPs  }
         \label{timing}
        \begin{tabular}{lcccccc}
            \hline
            \noalign{\smallskip}
  Observation & Satellite  & Flux$^{(a)}$      & N$_H$              & $\alpha_{ph}$ / kT$_{BB}$ & R$_{BB}^{(b)}$ & References    \\
  date       &          & (erg~cm$^{-2}$~s$^{-1}$) & ($10^{22}$~cm$^{-2}$) & (- /keV)     & (km) &     \\
            \noalign{\smallskip}
            \hline
\noalign{\smallskip}
\textbf{\ee}       &     & &      & & &    \\
\noalign{\smallskip}
3-4 May     1993        & \textit{ASCA}    & $3.2\times10^{-11}$  & 0.45 & 3.2 / 0.42   & 3$d_4$ & Corbet et al. (1995) \\
1,11 August 1995       & \textit{ASCA}    & $2.5\times10^{-11}$  &  0.52 & 3.6 / 0.41 & 3.1$d_4$ & This work    \\
16-17 November  1996      & \textit{SAX}     & $2.9\times10^{-11}$  & 0.87  &  3.93 / 0.44 & $3.3d_4$ & Parmar et al. (1998) \\
11 January  2000       & \textit{Chandra}  & $3.1\times10^{-11}$  & 0.93 & 3.6 / 0.41 &  $4.4d_4$ & Patel et al. (2001) \\
\hline
\noalign{\smallskip}
\textbf{\oo}       &     & &       & & &   \\
\noalign{\smallskip}
2,7 March  1994        & \textit{ASCA}     & $7.0\times10^{-12}$  & 1.0 & 2.9 / 0.57 & 0.96$d_5$ & Paul et al.  (2000) \\
10-11 May  1997        & \textit{SAX}     & $8.3\times10^{-12}$  & 1.2 & 3.3 / 0.62 & 0.97$d_5$ &  Tiengo et al. (2002) \\
26-27 July 1998       & \textit{ASCA}     & $7.0\times10^{-12}$  & 1.2  & 3.2 / 0.56 & 1.09$d_5$ &  Paul et al.  (2000) \\
28 December 2000        & \textit{XMM}     & $5.7\times10^{-12}$  & 1.0 & 2.9 / 0.63 & 0.74$d_5$ &  Tiengo et al. (2002) \\
\hline
\noalign{\smallskip}
\textbf{\uu}        &     & &   & & &       \\
\noalign{\smallskip}
18-19 September  1998     & \textit{ASCA}     & $13.3\times10^{-11}$  & 1.10 & 3.84 / 0.38& 2.3$d_1$ &  Paul et al.  (2000) \\
3 January 1997        & \textit{SAX}     & $9.9\times10^{-11}$  &  1.11 & 3.86 / 0.42 & 1.5$d_1$ & Israel et al. (1999a) \\
3 February 1998       & \textit{SAX}     & $9.5\times10^{-11}$  &  0.98 & 3.58 / 0.36 & 2.1$d_1$ & Israel et al. (1999a) \\
21 August   1998       & \textit{ASCA}     & $11.6\times10^{-11}$  &  1.17 & 3.87 / 0.38 & 2.2$d_1$ &  Paul et al. (2000) \\
July/August 1999     & \textit{ASCA}     & $9.3\times10^{-11}$  & 1.04 & 3.92 / 0.40 & 1.8$d_1$ &  This work \\
23 May 2001     & \textit{Chandra}     & $12.1\times10^{-11}$  & 0.88 & 3.3 / 0.42 & 2.0$d_1$ &  Juett et al. (2002) \\
\hline
\noalign{\smallskip}
\textbf{\rxs}        &     & &     & & &     \\
\noalign{\smallskip}
3 September 1996    & \textit{ASCA}     & $4.4\times10^{-11}$ & 1.42 & 2.92 / 0.41 & 7.1$d_8$  &  Sugizaki et al. (1997)  \\
1 April 1999        & \textit{SAX}     & $4.5\times10^{-11}$  & 1.42 & 2.62 / 0.46 & 6.5$d_8$  &  Israel et al. (2001) \\
September  1999     & \textit{ASCA}    &  $2.9\times10^{-11}$  & 1.49 & 3.1 / 0.45 & 5.1$d_8$  &  This work          \\
\hline
\noalign{\smallskip}
\textbf{\kes} & & &  & & & \\
\noalign{\smallskip}
11-12 October 1993  & \textit{ASCA} & $1.3\times10^{-11}$ & 2.0 & 3.0 / -- & -- & Vasisht \& Gotthelf (1997)   \\
27 March 1998    & \textit{ASCA}   & $1.1\times10^{-11}$     & 2.21 & 3.38 / -- & -- &  This work  \\
March/April 1999  & \textit{ASCA} &  $1.2\times10^{-11}$  & 2.26 & 3.46 / -- & -- &  This work          \\
\hline
\noalign{\smallskip}
\textbf{\axj}        &     & &   & & &        \\
\noalign{\smallskip}
12 October     1993        & \textit{ASCA}     & $3.9\times10^{-12}$  &  4.6 & -- / 0.72 & 1.5$d_8$ & Torii et al.  (1998) \\
12 October      1993        & \textit{ASCA}     & $4.3\times10^{-12}$  &   9.0 & 4.6 / -- & -- & Torii et al.  (1998) \\
28-29 March     1999        & \textit{ASCA}     & $\lsim4\times10^{-13}$  &--  & --&-- &  Vasisht et al. (2000) \\
\hline
        \end{tabular}
\begin{list}{}{}
\item[$^{\rm (a)}$] Flux in the 1-10 keV range, not corrected for the absorption
\item[$^{\rm (b)}$] $d_n$ is the assumed distance in units of $n$ kpc
\end{list}
   \end{table*}

\subsection{Long term variability}

Most AXPs have   been detected at flux levels consistent with
a constant luminosity. However the observations were carried out
with different instruments and the
limits that one can infer on the absence of variability are subject to systematic
uncertainties. Besides the unavoidable (model dependent) uncertainty
deriving from conversions between different energy ranges, the fluxes obtained
with non-imaging instruments
must also be corrected for the (poorly known)
contribution of other components in the field of view
(e.g. SNRs, diffuse galactic ridge emission, confusing sources, etc...).

Keeping this  in mind, we summarize in Table 2 all
the measurements obtained with imaging
instruments in the $\sim$ 1--10 keV range.
Although small differences in the best fit spectral parameters, in particular
N$_H$ and $\alpha_{ph}$, can result in large   variations if one considers
the \textit{unabsorbed} luminosity in a range extending to lower energies,
no strong variability is shown
by the values of the \textit{observed} 1-10 keV flux reported in Table 2.

Comparison with some of the data  obtained before 1993  (and not reported in Table 2) requires
some caution. For instance, the \textit{EXOSAT} data of \uu\ were clearly contaminated
above a few keV by the nearby pulsar RX J0146.9+6121 (Mereghetti et al. 1993);
at lower energies
they seem to be consistent with the values of Table 2.
The flux measured with  \textit{Einstein} and \textit{ROSAT} for \kes\ was found to be constant by
Helfand et al. (1994).
Several observations exist for \oo\ and \ee, and for both sources
evidence for flux variations
obtained by comparing data from the same instrument were reported.
According to Seward et al. (1986), \oo\ was at least a factor 10 fainter in December 1978  than
at the time of its discovery six months later.
Both observations were obtained with the IPC detector on the
\textit{Einstein} satellite, but during the first one
\oo\ lied very close to the edge of the field of view and was not detected.
A more convincing case for long term variability is given by   \textit{GINGA} observations
of \ee\ (Iwasawa et al. 1992).
In the August 1990  observation the source
was  a factor $\sim$2  brighter than   in previous  measurements
(June 1987 and December 1989).
During the higher intensity state the source also showed a different pulse shape
(a larger difference in the relative intensity of the two pulses),
as well as a variation in the spin-down rate.

Large flux variability was seen in
the AXP candidate  \axj. This source  had   a flux
of 4.2$\times$10$^{-12}$ erg cm$^{-2}$ s$^{-1}$ (2-10 keV)  during an
\textit{ASCA} pointing carried out in December 1993, but was not visible
3.5 years later, implying  a flux decrease greater than a factor 14
(Torii et al. 1998).
A further \textit{ASCA} observation (March 1999)  revealed only a weak source
(F$_x$ $\sim$4$\times$ 10$^{-13}$ erg cm$^{-2}$ s$^{-1}$)
at a position consistent with that of the AXP (Vasisht et al. 2000).
A search for pulsations could not
be performed, due to the small number of counts, thus it is not clear whether
this source is really \axj\ in a low state.
The possible existence of    transient AXPs with low quiescent
luminosities has important implications for
the total number of AXPs in the Galaxy and   their inferred birthrate.

The level of variability in AXPs is of
interest since  it is expected that some emission processes
(e.g. thermal emission from the neutron star surface) produce less variability
than others (e.g. those involving mass accretion, which is  generally
subject to   fluctuations).
Detailed searches for correlations between luminosity   and spin-down
variations   could provide crucial evidence in favor of  accretion, whereby
different mass accretion rates  are expected to produce different torques on the
rotating neutron star.
For this reason a     regular monitoring of AXPs  with the \textit{RXTE} satellite has
been carried out for a few years. These observations allow to obtain phase connected
timing solution (Section 6.2) and to monitor intensity
changes. Unfortunately, due to the non-imaging nature of the PCA instrument and to
the uncertain background intensity variations, such data provide a reliable measure
only of the pulsed flux component.
For \uu, \ee, and \rxs\ upper limits (1$\sigma$) of 20-30\% on variations of the
pulsed flux have been derived (Gavriil \& Kaspi 2002).
No variations have been seen in \kes\ (Gotthelf et al. 2002), while the values of
pulsed flux measured in \oo\ over a 4 years time span have an \textsl{rms} at a
level of $\sim$30\% of the average, suggesting that this AXP is somewhat
less stable than the other ones (Kaspi et al. 2001).

\section{Timing properties}

\subsection{Folded light curves}

The folded light curves of all AXPs, derived from \textit{ASCA} GIS data,
are shown in Fig.2. They are characterized by broad pulses and rather smooth profiles.
The pulsed fractions range from about 7\%  for \uu\ and \kes\ up  to $\sim$70\% for \oo\ and do not depend
much on the energy. Two sources, \ee\ and \uu, clearly display  a double-peaked profile.
The light curve of \rxs\ can be interpreted as the sum of two pulses with different energy
spectra (see also Israel et al. 2001, Gavriil \& Kaspi 2002).

Large pulsed fractions are  generally not expected if the radiation is emitted close to
the surface of a neutron star, due to the gravitational light bending.
The effect of interstellar absorption is to increase the observed pulsed fractions
by preferentially absorbing the softest photons, which for a
thermally emitting neutron star have a lower modulation (Page 1995, Perna et al. 2000),
but it is not clear that this effect is effective enough for the
temperature and N$_H$ values typical of   AXPs.
According to De Deo et al. (2001),  a pulsed fraction
as high as that of \oo\ can be obtained only in the case of a strong beaming,
a requirement probably more easily provided by an accreting neutron star.

However, \"{O}zel et al. (2001) showed that models of emission from highly magnetized
atmospheres can account for the AXPs pulsed fractions, provided that the emission comes
from a single hot region on the neutron star surface.
More recently  \"{O}zel (2002) calculated the expected pulse profiles
for a variety of model parameters and concluded that magnetars with a single hot
emitting region give a better representation of the observations than models involving
emission from two antipodal regions.

\begin{figure*}
\begin{minipage}{6cm}
\centerline{\psfig{file=1e2259.ps,height=8.5cm} }
\end{minipage}
\begin{minipage}{6cm}
\centerline{\psfig{file=1e1048.ps,height=8.5cm} }
\end{minipage}
\begin{minipage}{6cm}
\centerline{\psfig{file=4u0142.ps,height=8.5cm} }
\end{minipage}
\hfill
\end{figure*}

\begin{figure*}
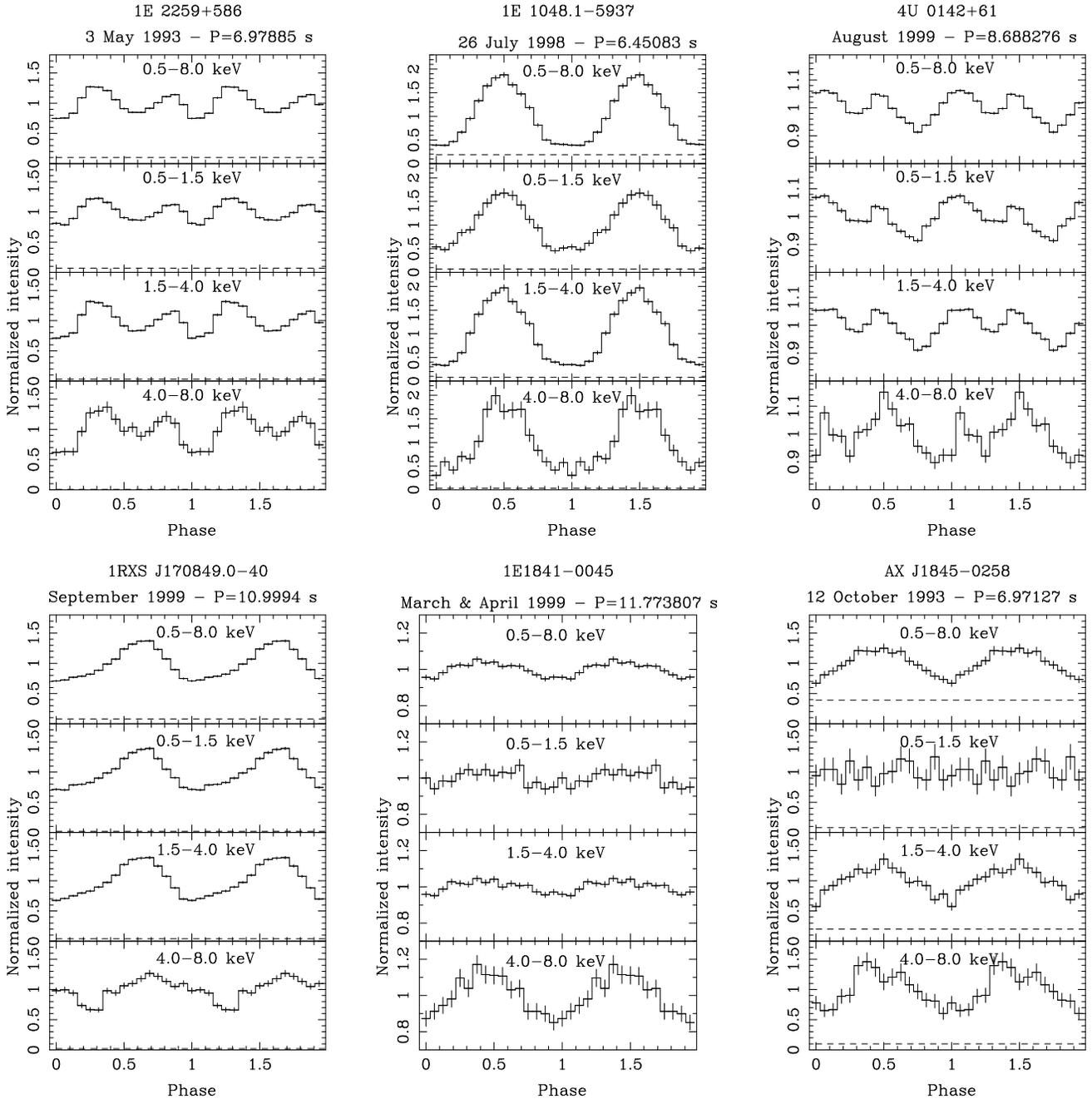

\begin{minipage}{6cm}
\centerline{\psfig{file=rxsj1708.ps,height=8.5cm} }
\end{minipage}
\begin{minipage}{6cm}
\centerline{\psfig{file=1e1841.ps,height=8.5cm} }
\end{minipage}
\begin{minipage}{6cm}
\centerline{\psfig{file=axj1845.ps,height=8.5cm} }
\end{minipage}
\caption{Folded light curves of the AXPs obtained with the \textit{ASCA} GIS instrument.
The dashed lines indicate the level of the background.}
\end{figure*}

\subsection{Secular spin-down and its fluctuations}

The  nearly steady spin-down of AXPs, in analogy with radio
pulsars,  immediately leads to consider their
characteristic ages defined as $\tau_c=P/2\pdot$ (see Table 1).
For a pulsar slowing down according to a braking law of the type
$\dot\nu\propto-\nu^n$  (or equivalently $\pdot\propto P^{2-n}$),
the characteristic age gives an upper limit on the true age
for $n$=3. The true age, given by

$${P\over(n-1)\dot P}\left[1-\left(P_0\over P\right)^{n-1}\right]$$

\noindent
can be significantly smaller than $\tau_c$ if the initial period $P_0$ is
close to $P$ and/or if $n$ is greater than 3.

Both \kes\ and \ee\  have  $\tau_c$  greater than the ages estimated for
the respective SNRs
\footnote{in the proposed association between SNRs and
SGRs we have the opposite situation, i.e. the $\tau_c$ values ($\sim$1500 yrs)
are smaller than the SNRs ages.}.
While in the first case the difference is relatively small,
for \ee, $\tau_c$ is at least ten times larger than the age of G109.1--0.1.
A possibility to explain this large discrepancy is to assume that
the pulsar suffered a much stronger braking in the past, before the magnetic
field decayed to a currently lower value (see, e.g., Colpi et al. 2000).

In principle, the study of   $\pdot$ fluctuations could
discriminate between accretion-based and magnetar models.
If accretion is at work
one would expect significant $\pdot$ variations,
possibly correlated with luminosity changes.
On the other hand, the period evolution of a magnetar should be much more
regular, with the possible exception of glitches, as observed in
young radio pulsars (see also
Melatos (1999) for magnetars $\pdot$ variations caused by   precession
driven by magnetic dipole radiation torques).
Until recently the few available period measurements did not allow  to discriminate
against alternative possibilities (see, e.g., Baykal \& Swank 1996, Heyl \& Hernquist 1999).
The extensive monitoring now provided by the \textit{RXTE} satellite
is producing more accurate data
in which phase-connected timing solutions can be obtained.
For \ee, these data show a very low level of timing noise during the last few years
(Kaspi et al. 1999), contrary to previous results that were based on sparse
observations spanning $\sim$20 years.
Also \rx\  was found to have a similarly low level of timing noise.
Both sources had phase residuals of only $\sim$1\%, comparable to or smaller
than those measured for most radio pulsars.
A sudden spin-up event was detected  in \rx\ in September 1999 and interpreted as a glitch
(Kaspi et al. 2000), thus reinforcing the similarity with young radio pulsars.

Another very stable rotator is \kes\ (Gotthelf et al. 2002), while \oo\
is characterized by a higher level of timing noise that prevented to find
a phase--coherent timing solution for time periods longer than a few months
(Kaspi et al. 2001).

The interpretation of this wealth  of new data is not straightforward.
First, the presence of glitches cannot  simply be taken as evidence for the
magnetar model: they are expected to occur in any neutron star subject to a
braking effect (although their detection in accreting systems
is hampered by the high level of timing noise).
Second, the HMXRB pulsar   4U\,1907+09
(P=440\,s), which  has spun--down constantly since  1983 at a rate only
a factor four greater than that of \ee,
has a long--term noise strength one order of magnitude lower than \ee\
(Baykal et al. 2000).
Another accreting pulsar with small timing noise is 4U\,1626--67 (Chakrabarty et al. 1997).

It seems therefore that further precise timing observations and more detailed model
predictions are required to distinguish between the various possibilities.

\subsection{Limits on orbital motion}

The most sensitive searches for orbital Doppler shifts  have been
carried out with  \textit{RXTE},
yielding  the upper limits on the projected  semi-major axis a$_x$sin{\it i}
reported in Table 3.
No sensitive searches for orbital
Doppler shifts have been performed for the other AXPs.
No periodic intensity variations, like eclipses or dips, that might indicate the presence
of a binary system, have been detected in any  AXPs.


\begin{table}
      \caption{Results of orbital motion searches}
        \begin{tabular}{lccc}
            \hline
            \noalign{\smallskip}
  SOURCE        &  a$_x$sin{\it i}  & Range  & References    \\
                  & (light-s)  &     of P$_{orb}$          &      \\
\noalign{\smallskip}
\hline
\noalign{\smallskip}
\ee      & $<$0.03   & 194 s - 1.4 d & (1)  \\
\ee      & $<$0.07   & 170 s - 5000 s & (2)  \\
\oo      & $<$0.06  & 200 s - 1.5 d & (1)  \\
\uu      & $<$0.26  & 70 s - 2.5 d & (3) \\
\hline
\noalign{\smallskip}
        \end{tabular}

(1) Mereghetti et al. (1998);
(2) Patel et al. (2001);
(3)  Wilson et al. (1999).
\end{table}


Although the lack of detectable Doppler shifts seems to support the
isolated nature of AXPs, it is worth noting  that  pulse arrival time
delays have remained undetected also in 4U\,1626--674, which  is  in a close binary system
(P$_{orb}$ = 42 min) with  a
very low--mass companion star.
For any assumed inclination angle and orbital period
the limits on a$_x$sin{\it i} constrain the possible companion mass.
As discussed in Mereghetti et al. (1998), late main sequence  stars are extremely
unlikely, but  He stars with mass  M$\lsim$ 0.8 $\msun$ and
white dwarf companions cannot be excluded.

\subsection{Period clustering}

The clustering of the AXPs periods in the narrow range 6--12 s is remarkable,
especially when compared with the large period distribution of pulsars in X--ray
binaries.
The absence of shorter period AXPs is not surprising given the small sample of
AXPs currently known and their secular spin-down.
Models, however, should be able to explain why no AXPs with longer periods are observed.

In accretion-based models, AXPs are supposed to rotate at (or close to)
the equilibrium period:

$$ P_{eq} \sim 2.5 B_{11}^{6/7} L_{35}^{-3/7} ~~~~\textrm{s}$$

\noindent
The observed period distribution and X--ray luminosities  can be obtained for neutron stars with magnetic
fields below the magnetar range and suitable values of the mass accretion rate
(Mereghetti \& Stella 1995).
It is unclear why only the   combination of these two parameters producing
periods in the 6-12 s range should be observed.
In models based on a fossil disk, in which the equilibrium period increases with the
secular decline of the accretion rate, some mechanism must be invoked to
turn off the AXPs at periods $\gsim$12 s.
Chatterjee et al. (2000) propose that an advection-dominated accretion flow
ensues when the accretion rate   decreases below a critical value,
thus limiting the active lifetime of AXPs to  $\gsim$5 10$^4$ yrs.

In  the context of the magnetar
scenario, the period distribution can be explained   if
the magnetic field decays on a timescale of $\sim$10$^4$ years.
Models without a significant field decay would lead to
the presence of AXPs with longer periods (Colpi et al. 2000).

\section{Multi wavelength observations}

\subsection{Optical and Infrared counterparts}

AXPs lie in the galactic plane, where crowding and severe interstellar
absorption complicate the identification work.  In the lack of  localizations at the arcsecond level,
studies based on deep multi-color  photometry and spectroscopy of the brightest
candidates could only establish the high X--ray-to-optical flux ratio of AXPs, thus
ruling out the presence of massive counterparts.
A typical situation is shown in Fig. 3. The current observational situation is
summarized in Table 4.

The recently announced discovery of optical pulsations
at the X--ray period (Kern \& Martin 2001) has
clinched the identification
of \uu\ with a faint, relatively blue object first proposed by Hulleman et al. (2000).
A promising candidate has also been found for \ee, while only upper limits
have been derived for the other objects. These  upper limits
are less constraining in the AXPs with the highest absorption, such as \kes\ and \rx.
Much progress in the identification process is expected in the next few years thanks to the
more accurate positions that are becoming available  with the \textit{Chandra} satellite.

Both for \uu\ and   the proposed near IR counterpart of \ee\  the optical/IR
flux is higher than an extrapolation of the blackbody like component of
the X--ray spectrum, clearly indicating the need for a different origin.
An extended accretion disk around an isolated neutron star would probably produce
too much optical and IR emission compared to the observed fluxes
(Hulleman et al. 2000, Perna et al. 2000).
However, these predictions depend on several very uncertain parameters
(e.g. the inclination angle,    the inner and outer radius of the disk) and it is
not possible to definitely rule out the possibility of accretion, based on current data.
The data recently reported by Hulleman (2002) on the flux in the K and B bands for
\uu\ further complicate the interpretation of the optical/IR emission:
the K band flux lies above the extrapolation of the shorter wavelength points,
while in B only a very tight upper limit has been obtained, implying a spectral turn over.

  \begin{figure}
  \centerline{\psfig{file=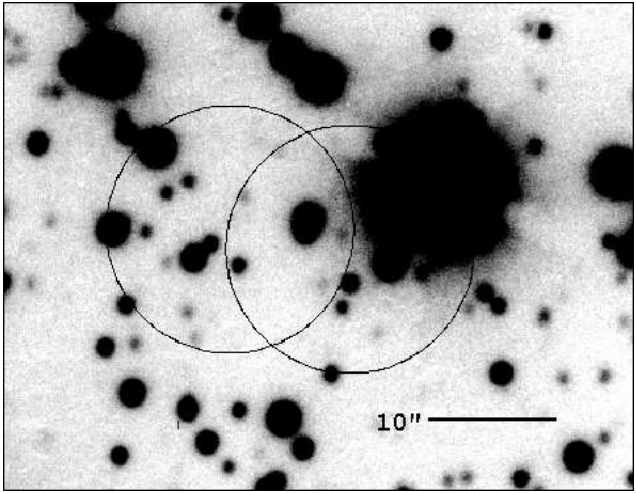,height=6.5cm} }
 \caption{R band image of the field of \rxs. The two circles indicate the  error regions obtained in two \textit{ROSAT} observations.
 \label{image}}
 \end{figure}

\begin{table*}
      \caption{Positions and optical counterparts}
         \label{timing}
        \begin{tabular}{lllccl}
            \hline
          \noalign{\smallskip}
  SOURCE    &   Coordinates   &   Galactic   &   Uncertainty   &   Optical &     Comments \\
 and references  &  (J2000)   &   Coordinates     &       &    & \\
           \noalign{\smallskip}
\hline
\noalign{\smallskip}
\ee    &   ~~23 01 08.295  &   ~~109.1  & 0.6$''$   & K=21.7$\pm$0.2 , R$>$26.4,     &  Possible IR counterpart         \\
  (1)      &  +58 52 44.45  &  -- 0.10  &   \textit{Chandra}    &  I$>$25.6, J$>$23.8       & at the \textit{Chandra} position   \\
\hline
\noalign{\smallskip}
\oo    &   ~~10 50 06.3  &   ~~288.3  & 4$''$   &  V$>$22     &   No stars with peculiar   \\
(2,3,4)  &  -- 59 53 17  &  -- 0.52 &  \textit{XMM-Newton}   &         & colors in the  error region \\
\hline
\noalign{\smallskip}
\uu    &   ~~01 46 22.41  &   ~~129.4 & --  &  V=25.6,  R=25      &  Pulsed optical counterpart         \\
(5,6,7)  &  +61 45 03.2   &  -- 0.43 &  optical  &  I=23.8,  K=19.6   &    \\
\hline
\noalign{\smallskip}
\rxs    &   ~~17 08 47.24  &   ~~346.5   & 10$''$  &  R$\gsim$18.2  &   Several  stars with R$\gsim$18    \\
 (8)    &  -- 40 08 50.7   &  +0.04 &  Rosat HRI $'$97 &         &   in  error region. None has  \\
\noalign{\smallskip}
     &   ~~17 08 46.52  &       & 10$''$  &          &     peculiar colors   \\
       &  -- 40 08 52.5   &    &  Rosat HRI  $'$93 &         &    \\
\hline
\noalign{\smallskip}
\kes   &   ~~18 41 19.0   &   ~~27.4  & 3$''$  &  R$\gsim$19     &  Several very reddened    \\
 (9,10)   &  -- 04 56 08.9  &   -- 0.005  &  \textit{Einstein} HRI        &  & stars in error region. Early    \\
\noalign{\smallskip}
 (11)   &   ~~18 41 19.2 &      & 3$''$  &       &  types cannot be excluded  \\
       &  -- 04 56 12.5  &     & Rosat HRI         &         &  if A$_{v}\gsim$8  \\
\hline
\noalign{\smallskip}
\axj    &   ~~18 44 57   &   ~~29.5 & 120$''$  &       &    No published optical       \\
 (12)       &  -- 03 00   &  +0.07   &   \textit{ASCA}       &         &  observations  \\
\noalign{\smallskip}
  (13)     &   ~~18  44 53  &       & 20$''$  &       &    Position of  \\
        &  -- 02 56 40   &   &  \textit{ASCA}   &         & AX J184453-025640, a faint \\
        &    &     &     &   &     source in the error box of \\
        &    &     &     &   &    the pulsar  \\
            \noalign{\smallskip}
             \hline
            \noalign{\smallskip}
        \end{tabular}
(1) Hulleman et al. (2001);
(2) Tiengo et al. (2002);
(3) Mereghetti et al. (1992);
(4) Wang \& Chakrabarty (2002);
(5) Hulleman et al. (2000);
(6) Kern \& Martin (2001);
(7) Hulleman (2002);
(8) Israel et al. (1999b);
(9) Kriss et al. (1985);
(10) Mereghetti et al. (2001);
(11) Helfand et al. (1994);
(12) Torii et al. (1998);
(13) Vasisht et al. (2000).
   \end{table*}

\subsection{(Lack of ?) radio emission}

A few searches for  radio emission from AXPs have been carried
out, some of which concentrated on revealing radio pulsations at the
X--ray period. All of them gave negative
results. We    summarize the upper limits in Table 5.
The lack of radio emission is not unexpected if AXPs are powered by accretion;
different explanations have been put forward
in the context of the magnetar model. An obvious possibility is
to invoke an unfavorable beaming angle, also considering that   radio
pulsars show an anti-correlation  between the beam width and the  rotation period.
Another explanation might be related to the fact that in a high magnetic field
(B$\gsim10^{14}$ G) the process of photon splitting becomes competitive
with  pair creation, thus suppressing the magnetospheric
cascades that are at the origin of the radio emission (Baring \& Harding 1998, 2001).
Gaensler et al. (2001) pointed out, however,  that the
current   upper limits are not very constraining: many radio  pulsars have 400 MHz
luminosities below those inferred from the AXPs  upper limits.

\begin{table}
\begin{center}
      \caption{Upper limits on  radio emission at 1.4 GHz}
         \label{timing}
        \begin{tabular}{lcl}
            \hline
            \noalign{\smallskip}
  SOURCE        & 5$\sigma$ u.l. & References   \\
           & (mJy) &    \\
  \noalign{\smallskip}
      \hline
            \noalign{\smallskip}
\ee        & 0.083       & Coe et al. (1994) \\
\noalign{\smallskip}
\oo      & 0.07$^{(a)}$  &   Israel et al. (2002)       \\
         & 0.13$^{(a)}$  &   Crawford et al. (2002) \\
\noalign{\smallskip}
\uu      & 0.27          &     Gaensler et al. (2001)\\
\noalign{\smallskip}
\rxs     &  3            &    Gaensler et al. (2001) \\
         & 0.07$^{(a)}$  &  Israel et al. (2002)  \\
         & 0.22$^{(a)}$  &  Crawford et al. (2002)      \\
\noalign{\smallskip}
\kes    &   0.6          &  Kriss et al. (1985)   \\
        & 0.07$^{(a)}$  &  Israel et al. (2002)    \\
        & 0.25$^{(a)}$ &  Crawford et al. (2002)  \\
\noalign{\smallskip}
\axj    & 0.07$^{(a)}$ &  Israel et al. (2002)    \\
        & 0.13$^{(a)}$ &   Crawford et al. (2002)   \\
            \noalign{\smallskip}
             \hline
\noalign{\smallskip}
        \end{tabular}
\begin{list}{}{}
\item[$^{\rm (a)}$] search for pulsed emission
\end{list}
\end{center}
\end{table}


\section{Relations with other classes of neutron stars and evolution}

\subsection{Association with Supernova Remnants}

Gaensler et al. (2001) have critically reexamined the evidence for SNRs
associated to AXPs and SGRs, concluding that  \ee\ and \kes\ are indeed the most
convincing associations.
These sources are located very close to the center of the two well studied
shell-like SNRs G109.1--0.1 (CTB 109) and Kes 73 (G27.4+0.0).
There is no need to invoke large velocity neutron stars, as in the case of   SGRs.
For the same reasons also  the association between \axj\ and G29.6+0.1, although these
objects have not been studied in detail, is quite convincing.

The presence of these SNRs is a very strong argument in favor
of an age smaller than a few $\times$10$^{4}$ years for the AXPs.
This conclusion is not contradicted by the apparent absence of a SNR around \uu\ and \oo,
since several radio pulsars with characteristic ages smaller than $\sim$10$^{4}$ years
do not have associated remnants.

\subsection{Soft Gamma-ray Repeaters}

Having     mentioned the similarities between AXPs and SGRs  (Section 4.2)
we concentrate here on some important differences between these two classes.
Quiescent SGRs have rather hard X--ray spectra, that, when fitted with power laws, give photon
index $\alpha_{ph}\sim$2 (see, e.g., Mereghetti et al. 2000).
Woods et al. (1999) found that the addition of a blackbody component
with kT$_{BB}$=0.5 keV resulted in  a better
fit to the X--ray spectrum of SGR 1900+14. Such a component was required only
during an observation in which no bursting activity was present, thus supporting the
interpretation of AXPs as inactive SGRs. We note however that the blackbody
plus power law spectrum of SGR 1900+14 was still  harder than the typical AXP spectrum
(see Fig.4).

SGRs also differ from AXPs for what concerns their location with respect to
the (possibly) associated SNRs, which requires  large transverse velocities
($\sim$1000 km s$^{-1}$). This might imply that AXPs
evolve into SGRs as they age and move away from the SNR centre
(Gaensler et al. 2001). A difference of 10--100\,kyr in the age
of AXPs and SGRs would be required in this case, thus making  it difficult to explain
the similarity of periods in the light
of the observed period derivatives.

\begin{figure}
\centerline{\psfig{file=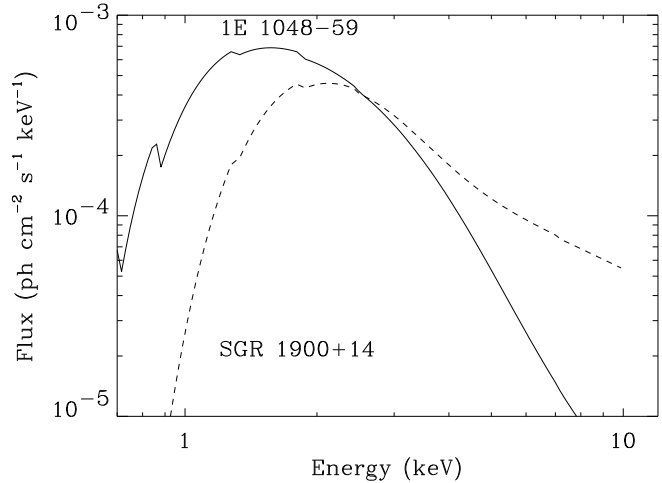,height=7.cm,angle=90} }
\caption{Comparison  of model X--ray spectra for an AXP
(solid line, \oo, Tiengo et al. (2002)) and a SGR
(dashed line, SGR 1900+14, Woods et al. (1999)). \oo\ is the AXP
with the hardest spectrum.
\label{image}}
\end{figure}

\begin{table*}
      \caption{Other sources possibly related to AXPs}
         \label{timing}
        \begin{tabular}{lcccccl}
\hline
\noalign{\smallskip}
\multicolumn{7}{c}{{\bf Soft Gamma-ray Repeaters}} \\
\noalign{\smallskip}
\hline
\noalign{\smallskip}
             \noalign{\smallskip}
  SOURCE   & $P$  & $\pdot$      & L$_{x}$       & $\alpha_{ph}$ & d     & Comments \\
           & (s)  & (s~s$^{-1}$)   &(erg s$^{-1}$) &      &  (kpc) & and references \\
            \noalign{\smallskip}
            \hline
            \noalign{\smallskip}
SGR 0525--66    &   8.1   & --                     & $\sim10^{36}$     &  2  & 55 &  in N49 in LMC (1) \\
SGR 1806--20    &   7.45  & $\sim$8.3$\times10^{-11}$ & $\sim4\times10^{35}$      &  1.95--2.25 & 15 & (1,4,5) \\
SGR 1900+14    &   5.16   &  $\sim$6.1$\times10^{-11}$ & $\sim2-8\times10^{34}$   & 1.9--2.2 & 5&  (1,6,7)   \\
SGR 1627--41    &   --   & --  & $\sim10^{35}$         & 2.5 & 11 &  (1,8)   \\

\hline
\noalign{\smallskip}
\multicolumn{7}{c}{{\bf Compact Central Objects in SNRs}} \\
\noalign{\smallskip}
\hline
\noalign{\smallskip}
  SOURCE   & $P$  & $\pdot$      & L$_{x}$       & kT$_{BB}$ & d     & Comments \\
           & (s)  & (s~s$^{-1}$)   &(erg s$^{-1}$) & (keV)     &  (kpc) & and references  \\
            \noalign{\smallskip}
            \hline
            \noalign{\smallskip}
RX~J0822--4300    &   --   & --  & $4.7\times10^{33}$         & 0.38 & 2 &  in  Puppis A  (2,9,10) \\
AX J0851.9--4617.4 &  --  & --   &  $9.1\times10^{32}$       & 0.4 & 2 & in  G266.1-1.2   (2,11,12) \\
1E~1207.4--5209    &  0.424   & $(2^{+1.1}_{-1.3})\times10^{-14}$   &  $9.5\times10^{32}$  & 0.27 & 2.1 & in  G296.5+10  (2,13,14)  \\
1E~161348--5055.1    &   --   &  --  &  $(2.1-100)\times10^{33}$       & 0.6 & 3.3 & in  RCW 103 (2,15)    \\
CXO J232327.8+584842  &   --  & --    &  $3.4\times10^{33}$    & 0.7 & 3.4 & in Cas A   (2,16,17,18) \\
\hline
\noalign{\smallskip}
\multicolumn{7}{c}{{\bf Dim Thermal Neutron Stars}} \\
\noalign{\smallskip}
\hline
\noalign{\smallskip}
%
RX~J0420.0--5022 &    22.69    & --    & $2.7\times10^{30}$     & 0.057 & 0.1 &  (3,19)  \\
RX~J0720.4--3125  &  8.39     & $(3-6)\times10^{-14}$  & $2.6\times10^{31}$    &  0.079       & 0.1  &   (3,20,21) \\
RX~J0806.4--4123   &   --    & -- &        $5.7\times10^{30}$ & 0.078   & 0.1 &  (3)  \\
RX~J1308.8+2127   &   5.16    &  $(1.35\pm0.7)\times10^{-11}$ &   $5.1\times10^{30}$ & 0.118   & 0.1 & RBS 1223 (3,22,23)  \\
RX~J1605.3+3249   &   --    & -- &        $1.1\times10^{31}$ & 0.092   & 0.1 & RBS 1556  (3,24) \\
RX~J1836.2+5925   &   --    & -- &        $5.4\times10^{30}$ & 0.043   & 0.4 &   (25) \\
RX J1856.5--3754   &   --      & -- &  $1.5\times10^{31}$  & 0.057 & 0.117 &   (3,26,27) \\
1RXS~J214303.7+065419   &  -- & -- &   $1.1\times10^{31}$ & 0.090   & 0.1 & RBS 1774  (28) \\
         \noalign{\smallskip}
             \hline
         \noalign{\smallskip}
      \end{tabular}
(1) Hurley (2000);
(2) Pavlov et al. (2002a);
(3) Treves et al. (2000);
(4) Mereghetti et al. (2000);
(5) Kouveliotou et al. (1998);
(6) Feroci et al. (2001);
(7) Hurley et al. (1999a);
(8) Hurley et al. (2000);
(9) Petre et al. (1996);
(10) Zavlin et al. (1999);
(11) Pavlov et al. (2001);
(12) Mereghetti (2001);
(13) Mereghetti et al. (1996);
(14) Pavlov et al. (2002b);
(15) Gotthelf et al. (1999)
(16) Murray et al. (2002);
(17) Chakrabarty et al. (2001);
(18) Mereghetti et al. (2002);
(19) Haberl et al. (1999);
(20) Haberl et al. (1997);
(21) Zane et al. (2002);
(22) Schwope et al. (1999);
(23) Hambaryan et al. (2002);
(24) Motch et al. (1999);
(25) Mirabal \& Halpern (2001);
(26) Pons et al. (2002);
(27) Walter \& Lattimer (2002);
(28) Zampieri et al. (2001).
\end{table*}


\subsection{Compact central objects in SNRs}

We list in Table 6 a few X--ray sources that are  located within SNRs
and have not been detected as radio pulsars.
These sources, recently named Compact Central Objects (CCOs, see
Pavlov et al. 2002a, for a review), have some similarities with AXPs.
It is very likely that all   the CCOs  are young neutron stars, as
indeed confirmed for the ones from which pulsations have been detected.
They have soft thermal spectra with temperatures
similar to those of the blackbody components of AXPs,
but their luminosity is on average smaller (see Fig.5).
More sensitive searches for pulsations might lead to classify some of the CCOs
as new AXPs (this is certainly not the case for the
CCO in G296.5+10.0 which has  a short period).

Similar to AXPs, the CCOs have a rather constant X--ray flux, with the interesting
exception of 1E~1614--5055, the CCO in RCW~103
(Gotthelf et al. 1999, Garmire et al. 2000).
This source, together with \axj, might represent  an object
with intermediate properties between  AXPs and CCOs.

\begin{figure*}
\centerline{\psfig{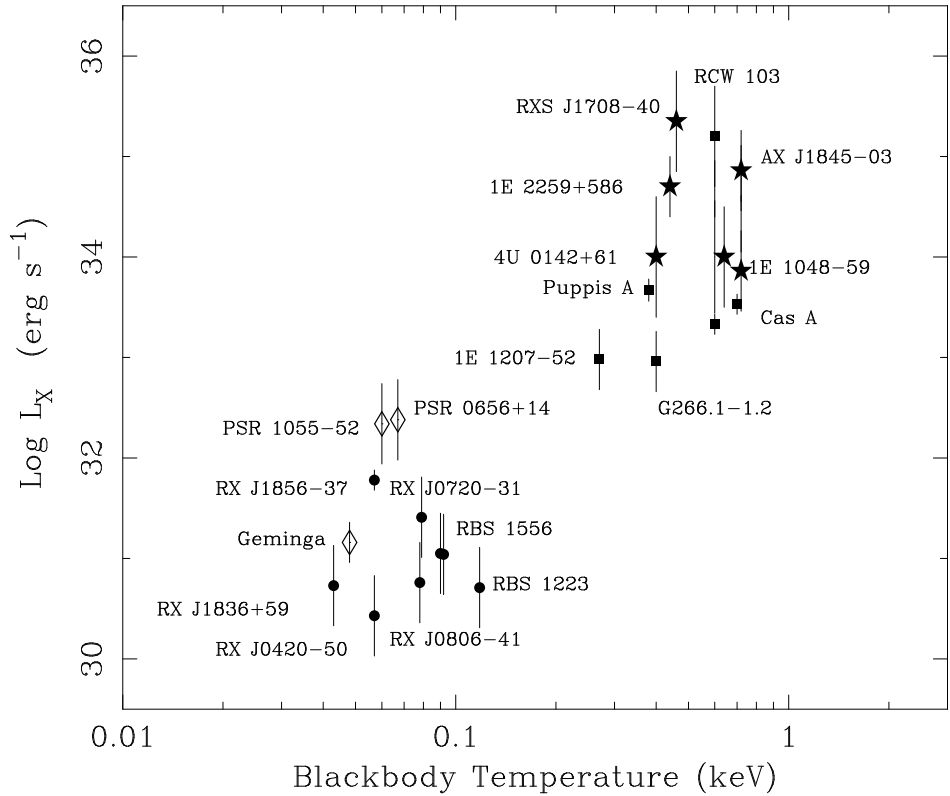} }
\caption{
''Color-Magnitude'' diagram for different classes of neutron stars: AXPs
(stars), CCOs in supernova remnants  (squares), and DTNs (circles).
For comparison also a few radio pulsars with thermal X--ray emission are shown (diamonds).
For the luminosity of AXPs and   radio
pulsars only the thermal spectral component has been  considered.
To facilitate the comparison across
different objects we plotted the temperatures inferred from pure blackbody fits; the use of
atmospheric models would move all the points to lower temperatures by a factor $\sim$2-3.
The luminosity error bars reflect the distance uncertainties. The two positions for the CCO
in  RCW~103 and for \axj\ indicate the observed variability range.
\label{image}}
\end{figure*}

\subsection{Dim thermal neutron stars}

Another class of neutron stars possibly related to the AXPs are the so called
Dim Thermal Neutron Stars (DTNs) discovered as very soft X--ray sources with the
ROSAT satellite (for a review see Treves et al. 2000).
Pulsations with periods similar to (or slightly longer than) those of AXPs have
been detected in four  of these sources. This confirms their neutron star nature
already suggested by their very high X--ray to optical flux ratio.
As shown in Fig. 5,  DTNs have softer spectra and lower luminosity than the CCOs and AXPs.
Their small distance also implies that they are a much more numerous and
probably older population. They are not associated with SNRs.
Both accretion from the ISM
and internal cooling have been considered to explain their X--ray emission.

Alpar (2001) proposed a unified scenario  in which the
DTNs, AXPs and CCOs are all interpreted as neutron stars with ordinary
magnetic fields (10$^{11}$-10$^{13}$ G).
According to Alpar (2001),  isolated neutron stars experience a mass inflow due to
the formation of a residual accretion disk from fallback material after the supernova explosion.
This causes an evolutionary phase in the propeller regime, before the neutron star
turns on as a radio pulsar or as an AXP, depending on the initial
rotational period, magnetic field and mass in the residual disk
(see also Chatterjee et al. 2000 for a similar model).
The DTNs in the lower left corner of Fig.5 are interpreted
as examples of sources in the propeller regime, in which the X--ray emission is
from cooling powered by internal friction. The CCOs would be the propeller objects
with the highest mass inflow.
According to this model they are surrounded by an optically thick corona and might evolve
into AXPs (Alpar 2001).

\section{Conclusions}

AXPs are one of the most enigmatic classes of
galactic X--ray sources.
Although the absence of a massive companion
and   presence of a neutron star
are observationally well established,
the current data are unable to discriminate between some of the most promising
models which have been proposed for these sources.

The fact that these models   involve very peculiar objects, the existence
of which is inferred from theoretical considerations,  but  not   yet  supported by compelling
observational data, is what makes the study of AXPs so exciting.

Much progress is expected in the next few years as deeper optical/IR and radio
observations will be carried out, together with more detailed studies with the \textit{Chandra},
\textit{XMM-Newton}  and \textit{RXTE} satellites. The X--ray data gathered in
the last few years are starting to reveal subtle differences among different AXPs.
These might provide crucial information to possibly relate AXPs with   other classes of
sources such as those mentioned in Section 8, in view of an overall understanding
of all the different manifestation of neutron stars.


{}

\end{document}